\magnification=1200
\settabs 18 \columns

\baselineskip=15 pt
\topinsert \vskip 1.25 in
\endinsert

\def\sqr#1#2{{\vcenter{\vbox{\hrule height.#2pt
 \hbox{\vrule width.#2pt height#1pt \kern#1pt
 \vrule width.#2pt} \hrule height.#2pt}}}}

\def\operp{\hbox{${\kern+.25em{\bigcirc}
\kern-.85em\bot\kern+.85em\kern-.25em}$}}

\def\lsim{\;\raise0.3ex\hbox{$<$\kern-0.75em\raise-1.1ex\hbox{$\sim$}}\;}
\def\gsim{\;\raise0.3ex\hbox{$>$\kern-0.75em\raise-1.1ex\hbox{$\sim$}}\;}
\def\no{\noindent}

\def\ce{\centerline}
\def\ve{\vfill\eject}
\def\rdots{\mathinner{\mkern1mu\raise1pt\vbox{\kern7pt\hbox{.}}\mkern2mu
 \raise4pt\hbox{.}\mkern2mu\raise7pt\hbox{.}\mkern1mu}}

\def\e e{$e^+ e^-$ }



\rightline{UCLA/03/TEP/34}
\rightline{November 2003}
\vskip1.5cm

\ce{\bf NOTE ON $q$-DUAL THEORIES}
\vskip.5cm

\ce{R. J. Finkelstein}
\vskip.3cm
\ce{\it Department of Physics and Astronomy}
\ce{\it University of California, Los Angeles, CA 90095-1547}
\vskip1.0cm

\no {\bf Abstract.}  The $q$-deformation of the Lie algebras underlying the
standard field theories leads to a pair of dual algebras.  We describe
a simple choice of possible field theories based on these derived algebras.  
One of these approximates the standard Lie theory of point particles, 
while the other is proposed as a field theory of knotted solitons.

\ve

\line{{\bf 1. Introduction.} \hfil}
\vskip.3cm

In discussing $q$-groups underlying non-standard field theories we shall use
the language of Lie groups rather than Hopf algebras, since we want to
emphasize the correspondence limit with standard theories by terming the
$q$-deformation of the standard Lie group an ``internal algebra" and  
the $q$-deformation of the standard Lie algebra an ``external
algebra".  Matrix elements of the external algebra are numerically valued and
are very close to the corresponding elements of the standard Lie algebra.
On the other hand, matrix elements of the internal algebra are not
numerically valued and have in general no finite matrix representation.

The external algebra, lying near the standard Lie algebra, may be made the
basis of a point particle field theory approximating the standard point
particle field theory.  The internal algebra, carrying degrees of freedom
present in neither the external algebra nor the corresponding point particle
theory, may be regarded as the basis of a soliton theory.  In the model here
proposed these new degrees of freedom, implying non-locality, will describe
solitons in the form of loops defined by the internal algebra.

We have previously discussed external $q$-gravity and $q$-electroweak and
found that these theories are very close to the standard theories: Einstein
gravity and Weinberg-Salam electroweak, but they are also incomplete.$^{1,2}$
Here we present a possible completion in the form of a solitonic version of
the internal theory.

\vskip.5cm

\line{{\bf 2. The Two Algebras.} \hfil}
\vskip.3cm

The two-dimensional representation of $SL_q(2)$ may be defined by
$$
T^t\epsilon_qT = T\epsilon_qT^t = \epsilon_q \eqno(2.1a)
$$
\no where
$$
\epsilon_q = \left(\matrix{0 & q_1^{1/2} \cr
-q^{1/2} & 0 \cr} \right) \qquad q_1 = q^{-1} \eqno(2.1b)
$$
\no Then if
$$
T = \left(\matrix{a & b \cr c & d \cr} \right) \eqno(2.2)
$$
\no eq. (2.1) implies
$$
\eqalign{ab &= qba \cr cd &= qdc \cr} \quad
\eqalign{ac &= qca \cr bd &= qdb \cr} \quad
\eqalign{bc &= cb \cr \hfil \cr} \quad
\eqalign{ad-qbc &=1 \cr da-q_1bc &= 1 \cr}\eqno(2.3)
$$
\no If $T$ is unitary,
$$
\eqalign{c &= -q_1\bar b \cr a &= \bar d} \eqno(2.4)
$$
\no The $2j+1$ dimensional representation of $SU_q(2)$ is
$$
\eqalign{D^j_{mm^\prime}(a,\bar a,b,\bar b) &= \Delta^j_{mm^\prime} \sum_{s,t}
\biggl\langle\matrix{n_+ \cr s\cr}\biggr\rangle_1
\biggl\langle\matrix{n_- \cr t\cr}\biggr\rangle_1 
q^{t(n_++1-s)}(-)^t \cr
&{}\times\delta(s+t,n_+^\prime)a^sb^{n_+-s}\bar b^t\bar a^{n-t} \cr} \eqno(2.5)
$$
\no where 
$$
\eqalign{n_\pm &= j\pm m \cr n_\pm^\prime &= j\pm m^\prime \cr} \qquad
\biggl\langle\matrix{n\cr s\cr}\biggr\rangle_1 = 
{\langle n \rangle_1!\over \langle s\rangle_1!
\langle n-s\rangle_1!} \qquad
\langle n\rangle_1 = {q_1^{2n}-1 \over q_1^2-1} 
$$
\no and
$$
\Delta^j_{mm^\prime} = \biggl[{\langle n_+^\prime\rangle_1!
\langle n_-^\prime\rangle_1!\over\langle n_+\rangle_1!
\langle n_-\rangle_1!}\biggr]^{1/2}
$$
\no Now set
$$
D^{1/2}(a,\bar a,b,\bar b) = e^{B\sigma_+}e^{\lambda\theta\sigma_3}
e^{C\sigma_-} \eqno(2.6)
$$
\no where $q=e^\lambda$ and expand to terms linear in $(B,C,\theta)$.  Then
$$
D^j_{mm^\prime}(a,\bar a,b,\bar b) = D^j_{mm^\prime}(0,0,0) +
B(J_B^j)_{mm^\prime} + C(J_C^j)_{mm^\prime} + 
2\lambda\theta(J_\theta^j)_{mm^\prime} + \ldots \eqno(2.7)
$$
\no The non-vanishing matrix coefficients $(J_B^j)_{mm^\prime}~
(J_C^j)_{mm^\prime}$ and $(J_\theta^j)_{mm^\prime}$ are by (2.5) and (2.7)
$$
\eqalign{\langle m-1|J_B^j|m\rangle &= \bigl[\langle j+m\rangle_1
\langle j-m+1\rangle_1\bigr]^{1/2} \cr
\langle m+1|J_C^j|m\rangle &= \bigl[\langle j-m\rangle_1
\langle j+m+1\rangle_1\bigr]^{1/2} \cr
\langle m|J_\theta^j|m\rangle &= m \cr} \eqno(2.8)
$$
\no Then $(B,C,\theta)$ and $(J_B,J_C,J_\theta)$ are the generators of the two dual
algebras referred to in the introduction.  They obey the following
commutation rules:
\vskip.3cm

\no {\bf External} algebra:
$$
(J_B,J_\theta) = -J_B \quad (J_C,J_\theta) = J_C \quad
(J_B,J_C) = q_1^{2J-1}[2J_\theta] \eqno(2.9)
$$
\no where
$$
[x] = {q^x-q_1^x \over q-q_1} \qquad
\langle x\rangle = {q^x-1\over q-1} 
$$
\no and
\vskip.3cm

\no {\bf Internal} algebra:
$$
(B,C) = 0 \quad (\theta,B) = B \quad (\theta,C) = C
\eqno(2.10)
$$
\no The matrix elements of the external algebra are given by (2.8) and (2.9).
They are numerically valued and describe a deformation of the $SU(2)$
algebra.

The arguments of $D^j_{mm^\prime}$, either $(B,C,\theta)$ or
$(a,\bar a,b,\bar b)$ are the generators of the internal algebra and describe
a deformation of the $SU(2)$ group.  They obey the algebra (2.10) or (2.3).
\vskip.5cm

\line{{\bf 3. The Dual Actions of $SU_q(2)$} \hfil}
\vskip.3cm

In the external theory the vector connection is expanded in the
external algebra:
$$
A_\mu(x) = \sum_{s=1}^3 A_\mu^s(x) J_s^1   \eqno(3.1)
$$
\no where $J_s^1$ are matrices whose numerically valued matrix elements satisfy (2.8)
with $j=1$.  The curvature is then
$$
\eqalign{F_{\mu\lambda} &= \sum\bigl[(\partial_\mu A_\lambda^s-\partial_\lambda
A_\mu^s)J_s^1 + A_\mu^sA_\lambda^t[J^1_s,J^1_t]\bigr] \cr
&= \sum(\partial_\mu A^p_\lambda-\partial_\lambda A^p_\mu +
f^p_{st}(q)A^s_\mu A^t_\lambda)J_p^1 \cr
&= \sum F^p_{\mu\lambda} J_p^1 \cr} \eqno(3.2)
$$
\no and the action is
$$
\eqalignno{S_{int} &= -{1\over 4} \int d^4x {\rm Tr}~F_{\mu\lambda}F^{\mu\lambda}
& (3.3) \cr
&= -{1\over 4} \int d^4x \sum F_{\mu\lambda}^p F^{\mu\lambda n} {\rm Tr}~
J^1_p J^1_n \cr
&= -{1\over 4} \int d^4x \sum g_{pn}(q) F^p_{\mu\lambda} F^{\mu\lambda n} &
(3.4) \cr}
$$
\no Here $f_{st}^p$ and $g_{pn}$ will depend on $q$ since the matrix
elements in (2.8) depend on $q$ if $j=1$.

In the internal theory the vector connection is expanded in the internal
algebra.
$$
A_\mu(a,\bar a,b,\bar b) = \sum_{jmn} A_\mu(x|jmn) D^j_{mn}(a,\bar a,b,\bar b)
\eqno(3.5)
$$
\no where $j$ is not restricted.  The curvature is now
$$
\eqalign{F_{\mu\lambda}(x|a,\bar a,b,\bar b) &= \sum_{jmn} \bigl[\partial_\mu
A_\lambda(x|jmn) - \partial_\lambda A_\mu(x|jmn)\bigr]
D_{jmn}(a,\bar a,b,\bar b) \cr
&{}~+ \sum_{\scriptstyle j_1m_1n_1\atop \scriptstyle j_2m_2n_2} A_\mu(x|j_1m_1n_1)A_\lambda(x|j_2m_2n_2) 
\bigl[D_{j_1m_1n_1}(a\bar ab\bar b),D_{j_2m_2n_2}(a\bar ab\bar b)\bigr]
 \cr} \eqno(3.6)
$$
\no To reduce the commutator expand the product of two $D_{jmn}$ as follows:
$$
D_{j_1m_1n_1}D_{j_2m_2n_2} = \sum_{jmn} \hat C_{j_1m_1n_1,j_2m_2n_2}^{jmn}
D_{jmn} \eqno(3.7)
$$
\no Then
$$
h(D_{j_1m_1n_1}D_{j_2m_2n_2}\bar D_{j_3m_3n_3}) = \sum_{jmn}
\hat C_{j_1m_1n_1,j_2,m_2,n_2}^{jmn} h(D_{jmn}\bar D_{j_3m_3n_3}) \eqno(3.8)
$$
\no where $h$ means a Woronowicz integration$^3$ over the $q$-group algebra
and
$$
h(D_{jmn}\bar D_{j_3m_3n_3}) = \delta_{jj_3}\delta_{mm_3}\delta_{nn_3}
{q^{-2n_3}\over [2j_3+1]_q} \eqno(3.9)
$$
\no Then
$$
\hat C_{j_1m_1n_1,j_2,m_2n_2}^{j_3m_3n_3} = q^{2n_3}
[2j_3+1]_qh(D_{j_1m_1n_1}D_{j_2m_2n_2}\bar D_{j_3m_3n_3}) \eqno(3.10)
$$
\no The triple integral over the $q$-group algebra is essentially a $q$-Clebsch-Gordan
coefficient.$^4$  Hence
$$
\eqalign{[D_{j_1m_1n_1},D_{j_2m_2n_2}] &= 
\sum_{j_3m_3n_3}(\hat C_{j_1m_1n_1,j_2m_2n_2}^{j_3m_3n_3} - \hat C_{j_2m_2n_2,j_1m_1n_1}^{j_3m_3n_3})
D_{j_3m_3n_3} \cr
&= \sum_{j_3m_3n_3} C_{j_1m_1n_1,j_2m_2n_2}^{j_3m_3n_3} D_{j_3m_3n_3} \cr} \eqno(3.11)
$$
\no where
$$
C_{12}^3 = \hat C^3_{12}-\hat C^3_{21} \eqno(3.12)
$$
\no Then
$$
\eqalign{
F_{\mu\lambda}(x|a\bar ab\bar b) &= \sum_{jmn}\bigl[\partial_\mu A_\lambda^{jmn}(x) -
\partial_\lambda A_\mu^{jmn}(x) \cr  
&~~+\sum_{\scriptstyle j_1m_1n_1\atop \scriptstyle j_2m_2n_2}C_{j_1m_1n_1,j_2m_2n_2}^{jmn}
A_\mu^{j_1m_1n_1}(x) A_\lambda^{j_2m_2n_2}(x)\bigr] D_{jmn} \cr}
\eqno(3.13)
$$
\no This is necessarily an infinite component field.  The sum in (3.13) is
restricted only by $j_1+j_2\geq j\geq |j_1-j_2|$.

The interior action is
$$
\eqalignno{S &= -{1\over 4} h\int d^4x F_{\mu\lambda}F^{\mu\lambda} & (3.14) \cr
&= -{1\over 4}\int d^4x F_{\mu\lambda}^{jmn} F^{\mu\lambda j^\prime m^\prime
n^\prime} h(D_{jmn}\bar D_{j^\prime m^\prime n^\prime}) \cr
&= -{1\over 4}\int d^4x \sum_{jmp} F_{\mu\lambda}^{jmp}
F^{\mu\lambda}_{jmp} {q^{-2p}\over [2j+1]_q} & (3.15) \cr}
$$
\no by (3.9).  Reduction of the action to a functional of $c$-number fields is accomplished
in the external action by tracing and in the internal action by $h$-integration.
To keep the correspondence with the external action we may restrict the
sum in (3.15) to $j=1$.  Then
$$
S_{int} = -{1\over 4} \int d^4x \sum_{1mp} F_{\mu\lambda}^{1mp}
F^{\mu\lambda}_{1mp} {q^{-2p}\over [3]_q} \eqno(3.16)
$$

The external and internal theories will differ in the two sets of structure
constants, namely: on the one side the external $f_{st}^p$ appearing
in (3.2) that are very close to the corresponding Lie structure constants
and on the other side the internal $C_{j_1m_1n_1,j_2m_2n_2}^{jmn}$ appearing
in (3.13) and dependent on the $q$-Clebsch-Gordan coefficients according to
(3.11), (3.12) and (3.10).  In addition, the curvature of the external
theory is linear in the generators $J_s$, 
while the curvature of the internal
theory is multinomial in the generators $(a,\bar a,b,\bar b)$.

Both $A^s_\mu(x)$ in (3.1) and $A_\mu(x|jmn)$ in (3.5) are to be Fock
expanded in states of momentum $(p)$ and polarization $(r)$.  Point
particles are to be associated with states $(p,r,s)$ and solitons with
states $(p,r,jmn)$.
\vskip.5cm

\line{{\bf 4. Interpretation of the Internal Theory.} \hfil}
\vskip.3cm

The external algebra lying near the standard Lie algebra has been
interpreted as the basis of a point particle field theory approximating
the standard point particle field theory.  The internal algebra carrying
degrees of freedom not present in the external algebra, nor in the
corresponding point particle theory, will now be interpreted as the basis of
a soliton theory.  One may also consider the possibility that the external
theory arising from linearization of (2.5) in (2.7) is a linearized version of the internal theory.
Since the internal $q$-theory is an infinite component theory, however, its
linearized version also has infinitely many components.  The external
$q$-theory, which is confined to the adjoint reprepresentation and has the
same number of components as the standard theory can therefore be related
to only a truncated form of the internal $q$-theory.

The interpretation of the internal theory depends on the algebra of
$SU_q(2)$.  We are assuming that the soliton states are labelled by the
irreducible representations of $SU_q(2)$ as in Eq. (3.5).  The $D^j_{mn}(a,\bar a,b,\bar b)$ appearing there are not numerically valued but are operators
whose expectation value may be computed on the state space defined by the
$(a,\bar a,b,\bar b)$ algebra.  Since $[b,\bar b] = 0$, choose common eigenstates of $b$ and $\bar b$ as the basic states.  Then $\bar a$ and $a$ are raising
and lowering operators.

Assume
$$
\eqalign{b|0\rangle &= \beta|0\rangle \cr
\bar b|0\rangle &= \beta^\star|0\rangle \cr}
$$
\no Then
$$
\bar b\cdot\bar a^N|0\rangle = q^N\beta^\star\cdot\bar a^N|0\rangle
$$
\no or
$$
\bar b|N\rangle = q^N\beta^\star|N\rangle
$$
\no There is an infinite number of states on which $\langle N|D^j_{mn}(a\bar ab\bar b)|N\rangle$ exists.$^5$  In this sense there is an infinite number
of degrees of freedom associated with each solitonic state $D^j(m,n)$.
These expectation values will be polynomials in $q,\beta$, and $\beta^\star$.
The states of excitation, $|N>$, of the soliton denoted by $D^j(m,n)$ are
analogous to the excited states of a string.  Here the eigenvalues of
$b\bar b$ are arranged in geometric rather than arithmetic progression.

The solitons may alternatively be characterized by Jones or Kauffman knot
polynomials that are defined by the $q$-algebra since $\epsilon_q$, the
$SU_q(2)$ invariant, encodes a program for these polynomials.  
To describe the Kauffman algorithm,$^6$ let $X$ be any crossing in the
graph $K$, representing an unoriented link, and let $\langle K\rangle$
be the polynomial associated with $K$:
$$
\langle K\rangle = \biggl\langle\matrix{\ldots \cr X \cr}\biggr\rangle \eqno(4.1)
$$
\no First express the polynomial $\langle K\rangle_n$ with $n$ crossings as
a linear combination of the polynomials $\langle K_-\rangle_{n-1}$ and
$\langle K_+\rangle_{n-1}$ each with $n-1$ crossings:
$$
\langle K\rangle_n = i~{\rm Tr}~\epsilon_q\bigl[\sigma_-\langle K_-\rangle_{n-1}
+ \sigma_+\langle K_+\rangle_{n-1}\bigr]  \qquad
\sigma_\pm = {1\over 2} (\sigma_1\pm i~\sigma_2) \eqno(4.2)
$$
\no where $K_-$ and $K_+$ signify two ways of splicing away the crossing
$X$ by opening up either the - or + channels.$^6$

Then iterating (4.2) to remove all crossings one obtains a linear combination
of $2^n$ polynomials and associated graphs, some with internal loops.  The
internal loops may then be removed by
$$
\eqalignno{\langle OK\rangle &= ({\rm Tr}~\epsilon_q\epsilon_q^t)\langle K\rangle & (4.3) \cr
\langle O\rangle &= {\rm Tr}~\epsilon_q\epsilon_q^t & (4.4) \cr}
$$

Application of these rules to a graph $K$ with multiple crossings reduces
$\langle K\rangle$ to a Laurent polynomial in $q$.  These rules relate
$\epsilon_q$, the invariant of $SU_q(2)$, to the polynomials characterizing
the unoriented knot.  If $K$ is oriented, one may form the following invariant of ambient isotopy$^6$
$$
f_k(A) = (-A^3)^{-W(K)}\langle K\rangle \eqno(4.5a)
$$
\no where
$$
A = i~{\rm Tr}~\epsilon_q\sigma_- \eqno(4.5b)
$$
\no and where $W(K)$, the sum of the crossing signs, is the writhe of $K$.  For
any oriented link
$$
V_k(t) = f_k(t^{1/4}) \eqno(4.6)
$$
\no is the one variable Jones polynomial.$^6$

It is then also possible to label a knot with the irreducible representation
$D^{N/2}_{{w\over 2}{r+1\over 2}}$ of $SU_q(2)$, where $N$, $w$, and $r$ signify
the number of crossings, the writhe, and the rotation (the Whitney degree
of the knot).  Here $N$, $w$, and $r$ are all integers and $w-r$ is
required to be odd by a knot constraint.  
We now assume some of the normal modes of the vector potential in (3.5) 
represent knots labelled by $D^{N/2}_{{w\over 2}{r+1\over 2}}$, while the
remaining normal modes represent either unknots (loops) or stringlike
structures with the excited states of a ``geometric" oscillator.

In the external $q$-electroweak theory the matrix elements in (2.8) 
and $J^1_s$ in (3.1) are labelled
by $(j,m)$ where $j$ is the isotopic spin $(I)$ and $m$ is its
$z$-component $(I_3)$.  Then the indices labelling $D^{N/2}_{{w\over 2}{r+1\over 2}}$ in
(2.7) and (3.5) must be related to the isotopic spin if the
external theory is regarded as a perturbative version of the internal theory,
i.e. $(N,w,r) \sim (I,I_3)$ and hypercharge.  

The eigenvalues of the Hamiltonian corresponding to (3.15) would then depend on these quantum
numbers as follows:  
$$
-{1\over 4}\int d\vec x \bigl[ F_{ok}(N,w,r)F_{ok}(N,w,r) +
F_{jk}(N,w,r) F_{jk}(N,w,r)\bigr]
{q^{-(r+1)}\over [N+1]}
$$
\no There is then an additional fine structure depending on $r$.

A knot labelled by $(N,w,r)$ and characterized by a Jones polynomial may be
termed a classical knot.  A knot labelled by $(N,w,r)$, but characterized by the operator \break $D^{N/2}_{{w\over 2}{r+1\over 2}}(a,\bar a,b,\bar b)$, defines a
state space and may be termed a quantum mechanical knot.  Both the polynomial and the state space stem from $SU_q(2)$.

The physical picture suggested here arises from the substitutiion of $SU_q(2)$ for $SU(2)$ in our standard theories.  New degrees of freedom are
necessarily introduced by the $q$-formalism and they are naturally
interpreted as solitonic.  The identification of these solitons as knots is then suggested by the relation of knots to the $SU_q(2)$
invariant $\epsilon_q$.  

Knot states appear in attempts to quantize general relativity.$^7$  Since
external $q$-gravity approximates standard gravity, we might expect knot
states to appear in external $q$-gravity as well.  Since the Kauffman rules
are based on $\epsilon_q$, the invariant of the $q$-Lorentz group, $SL_q(2)$,
the general argument of this section holds for internal $q$-gravity as well as 
for internal $q$-electroweak.  Therefore knots may appear in both internal
and external $q$-gravity.  Since all fields are coupled to the gravitational
field, knots may on these grounds be expected quite generally, and 
one may conjecture that$SU_q(2)$
plays the role of a universal hidden symmetry.

\vskip.5cm

\line{{\bf Ackowldgements.} \hfil}
\vskip.3cm

I would like to thank V. S. Varadarajan and A. C. Cadavid for helpful comments.

\vskip.5cm

\line{{\bf References.} \hfil}
\vskip.3cm

\item{1.} Finkelstein, R. J., Lett. Math. Phys. {\bf 38}, 53 (1996).
\item{2.} Finkelstein, R. J., Lett. Math. Phys. {\bf 62}, 199 (2002).
\item{3.} Woronowicz, S. L., RIMS, Kyoto {\bf 23}, 112 (1987).
\item{4.} Kirilov, A. N. and Reshitikhin, N. Yu, New Developments in
the Theory of Knots, World Scientific (1999); Cadavid, A. C. and
Finkelstein, R. J., J. Math. Phys. {\bf 36}, 1912 (1995).
\item{5.} Finkelstein, R. J., J. Math. Phys. {\bf 37}, 2628 (1996).
\item{6.} Kauffman, L. H., Int. J. Mod. Phys. A{\bf 5}, 93 (1990).
\item{7.} C. Rovelli and L. Smolin, Nuc. Phys. B{\bf 331}, 80 (1990).

\end
\bye